\documentclass{eas}
\usepackage{graphicx}
\usepackage{amssymb}
\usepackage{amsmath}
\renewcommand{\deg}{$^{\rm o}$}
\newcommand{\ion}[2]{#1\,{\sc #2}}
\renewcommand{\d}{\mathrm{d}}

\newcommand{\lya}{Ly$\alpha$}

\newcommand{\mearth}{M$_\oplus$}
\newcommand{\au}{{\sc au}}
\renewcommand{\etal}{et al.}
\TitreGlobal{Les Houches Winter School: Physics and Astrophysics of Planetary
Systems}
\begin{document}

%%-----------------------------
%%      the top matter
%%-----------------------------
\title{Evaporation of extrasolar planets}
\author{David Ehrenreich}\address{Laboratoire d'astrophysique de Grenoble\\ Universit\'e Joseph Fourier and CNRS (UMR 5571)\\ BP 53 38041 Grenoble cedex 9, France\\ david.ehrenreich@obs.ujf-grenoble.fr}
\begin{abstract}
Atomic hydrogen escaping from the extrasolar giant planet HD\,209458b provides
the largest observational signature ever detected for an extrasolar planet
atmosphere. In fact, the upper atmosphere of this planet is evaporating.
Observational evidences and interpretations coming from various models are
reviewed. Implications for exoplanetology are discussed.
\end{abstract}
\maketitle
%%-----------------------------
%%      your text
%%-----------------------------

\section{Introduction}
%=====================
\label{sec:intro} Extrasolar planets are now commonly found around main
sequence stars, usually by measuring the variations of the stellar velocity
projected on the line of sight. If the inclination of the planetary system is
close to 90\deg, the planet is seen from Earth passing in front of its host
star: this transiting event is the current best tool used to characterize
extrasolar planets. The transit of HD\,209458b, producing a $\sim 1\%$ dip in
the stellar light curve, has permitted to confirm the gaseous nature of
extrasolar giant planets (Charbonneau \etal\ 2000; Henry \etal\ 2000). Using
transmission spectroscopy, it was further possible to probe the composition and
structure of the planet atmosphere.

Transmission spectroscopy consists in measuring the depth of the transit light
curve -- which is related to the planetary radius -- as a function of
wavelength (see the contribution of Marley in this volume). Atmospheric sodium
is for instance detected as a supplementary absorption of $\sim 10^{-4}$ in the
\ion{Na}{i}~D lines at 589~nm (Charbonneau \etal\ 2002). Since the absorption
is located in the core of the lines, it appears stronger ($\lesssim 10^{-3}$)
at a better spectral resolution, thus probing higher altitudes in the planet
atmosphere (Sing \etal\ 2008; Snellen \etal\ 2008), typically $\sim 1\,000$~km
above the optically thick radius $r_p$ resulting from Rayleigh scattering by
molecular hydrogen (Lecavelier des Etangs \etal\ 2008). Yet, it remains a
tenuous signature in the visible.

On the other hand, intriguing strong absorptions were observed during the
planetary transit in some atomic lines located in the ultraviolet part of the
spectrum: the Lyman~$\alpha$ (\lya) line of neutral hydrogen (\ion{H}{i}) at
121.6~nm yields a dimming of 5 to 15\% depending on the spectral resolution
(Vidal-Madjar \etal\ 2003, 2004, 2008; Ben-Jaffel 2007; Ehrenreich \etal\ 2008)
and lines of atomic oxygen (\ion{O}{i}) around 130.5~nm and singly ionized
carbon (\ion{C}{ii}) close to 133.5~nm show a dip between 7 and 13\%
(Vidal-Madjar \etal\ 2004).

In fact, these absorptions are larger than the signature of the planet itself
during the transit. Hence, they must originate in the planet upper atmosphere
at the level of the Roche lobe. Consequently, this gas must be escaping the
planet gravity: HD\,209458b is evaporating. After giving some keys to the
understanding of the escape process (\S\ref{sec:defs}), I will review the
observational results (\S\ref{sec:obs}) and their possible interpretations
(\S\ref{sec:models}). Finally, I will discuss the implications and prospects of
evaporation in the frame of comparative exoplanetology
(\S\ref{sec:exoplaneto}).

\section{Exosphere and atmospheric escape: some definitions}
%===========================================================
\label{sec:defs} All the notions addressed in this chapter are developed by
Chamberlain \& Hunten (1987), where the reader is referred for more details.

\subsection{Collisionless atmospheres}
%-------------------------------------
\label{sec:collision} A planetary atmosphere can be roughly split in two parts:
the bottom part is the homosphere, where atmospheric constituants are mixed by
various processes (convection in the troposphere, eddy diffusion, molecular
diffusion, collisions, etc.) resulting in a more-or-less uniform composition
with a mean molar mass $\mu$. On the contrary, in the upper heterosphere, the
density is weak enough so that molecules and atoms are stratified by the planet
gravity as a function of their molar masses. The uppermost fringes of the
heterosphere constitute the exosphere.\footnote{Exo ($\acute{\varepsilon}\xi
o$) means `outside', just as in `exoplanet': a planet that lies outside of our
solar system. The exosphere is the boundary between an atmosphere and the space
outside.}

An exosphere is collisionless: the base of the exosphere, dubbed exobase or
critical level, is defined as the altitude $r_c$ where the atmospheric scale
height $H$ equals the mean free path $l$ of gas atoms (usually, hydrogen
atoms). Exospheres of close-in giant exoplanets extend up to their Roche radii
$r_R$ (Lecavelier des Etangs \etal\ 2004). In other words,
\begin{equation}
\int^{r_R}_{r_c} n(r) Q \d r \approx n(r_c) Q H = \frac{H}{l(r_c)} = 1,
\label{eq:collisionless}
\end{equation}
where $r$ is the distance to the planet centre, $n$ is the number density (in
g~cm$^{-3}$), Q is the collision cross section (cm$^2$), and $H =
\mathrm{R}T_c/\mu g_p$ with R (erg~K$^{-1}$~mol$^{-1}$) the universal gas
constant, $T_c$ (K) the exospheric temperature, and $\mu$ (g~mol$^{-1}$) the
mean molar mass of the atmosphere. The scale height can be considered constant
across the exosphere. An important point concerning giant exoplanet atmospheres
is that $T_c \gg T_\mathrm{eff}$.\footnote{For instance, on Jupiter $T_c
\lesssim 1\,000$~K, much higher than the planet effective temperature (Smith \&
Hunten 1990).}

\subsection{Escape processes}
%----------------------------
\label{sec:escape} Being a boundary layer between a relatively dense gas
envelope and space vacuum, the exosphere is where atmospheric mass loss
processes, or escape processes, take place. The simplest, yet usually not very
efficient, escape mechanism is the thermal escape (Jeans 1925). It assumes that
particles in the exosphere have a Maxwell-Boltzmann velocity distribution with
a most probable velocity $u$ depending on the temperature at the exobase as $u
= (2\mathrm{R}T_c/\mu)^{1/2}$. Only particles in the tail of the velocity
distribution, with speeds exceeding the escape velocity $v_\mathrm{esc} =
(2\mathrm{G}m_p/r_p)^{1/2}$ where G is the gravitational constant, can escape.
The Jeans escape flux in s$^{-1}$ is
\begin{equation}
F_J = \frac{n(r_c) u}{2\pi^{1/2}} (1 + \lambda_c) \exp{(-\lambda_c)},
\end{equation}
with the parameter $\lambda_c$ expressing the absolute value of the
gravitational potential energy at the exobase in units of $\mathrm{R}T_c$,
\begin{equation}
\lambda_c = \frac{\mathrm{G}m_p\mu}{\mathrm{R}T_c r_c} =
\left(\frac{v_\mathrm{esc}}{u}\right)^2 = \frac{r_c}{H}.
\end{equation}

At high $\lambda_c$, Jeans escape is a \emph{diffusion-limited} process because
the flux of escaping particles must be balanced by the flux of particles
reaching the exobase from below. It generally gives a lower limit to the actual
escape flux and remains valid providing $\lambda_c \gg 2$, i.e., the speed of
the expanding isothermal atmosphere is much less that the sound speed at the
exobase (Walker 1982).

The Jeans escape flux severely underestimate the measured escape rate of
planets in the solar system. Their atmospheric mass loss is in fact controlled
by `non-thermal' processes, a review of which can be found in Hunten (1982).
One mechanism known in the solar system and proposed to occur at HD\,209458b is
charge-exchange reactions between the neutral exosphere and stellar wind
protons (Holmstr\"om \etal\ 2008; see \S\ref{sec:models}), which can produce
energetic neutral atoms with high velocities, from the reaction $\rm H + H^{+*}
\rightarrow H^+ + H^*$.

Non-thermal escape processes are likely at play in the exospheres of extrasolar
planets. However, a major difference with the solar system is that $\sim 30\%$
of extrasolar planets has semi-major axis below 0.1 astronomical units (\au).
For these planets, the star is a tremendous source of energetic X and extreme
ultraviolet (EUV) photons that are able to heat the upper atmospheric layers
(the thermosphere) to $\sim 10^4$~K (Lammer \etal\ 2003). Ballester, Sing \&
Herbert (2007) provided an observational evidence of such high temperature by
detecting the signature of excited hydrogen atoms \ion{H}{i} ($n=2$), which
population peaks around $10^4$~K. Under these extreme heating conditions,
$\lambda_c$ reaches small values and the expanding atmosphere experiences a
`blow off' (Hunten 1982) where its bulk velocity becomes supersonic. At this
point, the best description of atmospheric escape is not that of Jeans, nor the
limit to the escape flux is the transport to the exobase; the description is
rather hydrodynamic and the energetic input to the upper atmosphere represents
the bottleneck to the escape flux.

Furthermore, close-in giant extrasolar planets are deeply embedded in the
gravitational well of their host stars, so that their extended atmospheres are
especially sensitive to tides. Lecavelier des Etangs \etal\ (2004) showed that
tidal forces enhance the escape rate of a hot Jupiter and pull the exobase
level up to the Roche limit, where the atmosphere is free to escape the planet
gravity, leading to a \emph{geometrical blow-off} (see also Gu, Lin \&
Bodenheimer 2003).

\subsection{Roche lobe}
\begin{figure}
\begin{center}
\includegraphics[width=0.5\columnwidth]{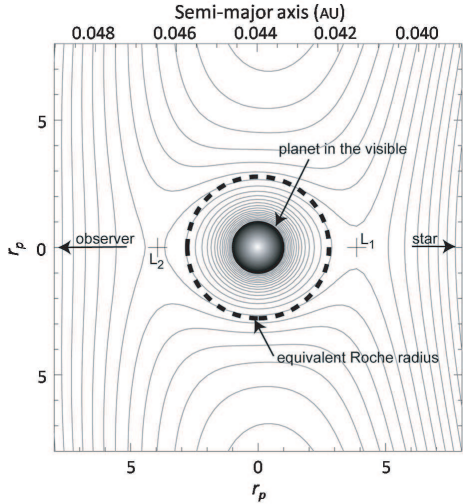}
\caption{\label{fig:roche} Gravitational equipotentials around HD\,209458b. The
Roche equivalent radius is delimited by the thick dashed circle. The actual
Roche limit is elongated toward the star and in the opposite direction. During
a transit, the observer see the small section of the Roche lobe, which size is
slightly overestimated by the equivalent Roche radius.}
\end{center}
\end{figure}

%----------------------
\label{sec:roche} The gravitational potential of a rotating circularized binary
system, such as a star-planet system, expresses (Hameury 2000)
\begin{equation}
\phi = -\frac{\mathrm{G}m_p}{|x-r_p|} - \frac{\mathrm{G}m_\star}{|x-r_\star|} -
\frac{1}{2} |\omega \times x|^2,
\end{equation}
where $x$ is the distance from the centre of mass and $\omega$ is the system
angular speed. The five points where the condition $\nabla \phi = 0$ is
satisfied are the Lagrangian points (L$_1$, \ldots, L$_5$), all located in the
orbital plane. The first 3 Lagrangian points are aligned with the centres of
the planet and the star. Figure~\ref{fig:roche} represents the equipotential
lines around the extrasolar planet HD\,209458b with the L$_1$ and L$_2$ points.
The Roche limit is the equipotential passing through L$_1$; it delimits the
lobe of the planet gravitational influence. The Roche lobe is elongated toward
the star and in the opposite direction and has therefore a radial
($r_\parallel$) and transverse ($r_\perp$) radii, with $r_\parallel > r_\perp$.
During a transit, the cross section of the Roche lobe is $4 \pi r_\perp^2$,
conveniently approximated by the equivalent Roche radius $r_R$, defined so that
$4/3\pi r_R^3$ is equal to the volume within the Roche lobe (Paczy\'nski 1971).
This radius can be estimated to better than 1\% by (Eggleton 1983)
\begin{equation}
r_\mathrm{Roche} = \frac{0.49q^{2/3}}{0.6q^{2/3} + \ln
\left(1+q^{1/3}\right)},\quad 0 < q < \infty,
\end{equation}
where $q = m_p / m_\star$ and the separation of the two body centres in unity.

\section{Observations of the exospheres of extrasolar planets}
%=============================================================
\label{sec:obs}

\subsection{Detection of the escaping exosphere of HD\,209458b from resolved data}
\begin{figure}
\begin{center}
\includegraphics[width=0.75\columnwidth]{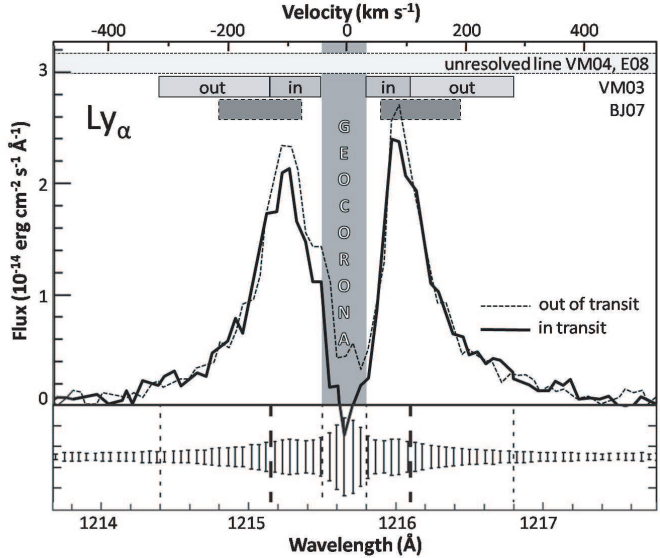}
\caption{\label{fig:lya} The \lya\ emission line of star HD\,209458. The
continuum is a double peaked emission originating from the stellar
chromosphere. The central part of the line is absorbed by interstellar
\ion{H}{i} and deuterium in the line of sight. It is also the place where the
geocoronal emission contaminates the spectrum. The 2003 STIS medium-resolution
spectrum is shown outside and inside the transit (upper panel) with the
$1$-$\sigma$ error bars (bottom panel). Bandpasses of the different \lya\
measurements are indicated, from top to bottom: Vidal-Madjar \etal\ (2004) and
Ehrenreich \etal\ (2008) determined the absorption by integrating over the
whole unresolved \lya\ line in 2004 STIS low-resolution and 2005 ACS
low-resolution data, respectively; Vidal-Madjar \etal\ (2003) normalized the
flux in the `in' domain by that in the `out' bands to correct for variations of
the stellar \lya\ flux (see Vidal-Madjar 1975 for a study of solar \lya\
variations); Ben-Jaffel (2007) integrated the flux of the 2003 STIS data in a
broader domain than Vidal-Madjar \etal\ (2003).}
\end{center}
\end{figure}

%---------------------------------------------------------------------------------
The transit of HD\,209458b was originally observed by Vidal-Madjar \etal\
(2003) with the Space Telescope Imaging Spectrograph (STIS; Woodgate \etal\
1998) on board the \emph{Hubble Space Telescope} (\emph{HST}) to search for
\ion{H}{i} in the planet atmosphere. Since \ion{H}{i} is the lightest possible
component of the planet, it could be detected high in the atmosphere, resulting
in a large absorption signal. Indeed, a surprisingly strong absorption of
$(15\pm4)$\% was measured in the \lya\ stellar emission line at 121.6~nm,
monitored during 3 transits and shown in Fig.~\ref{fig:lya}. The corresponding
light curve is plotted in the upper panel of Fig.~\ref{fig:light}. A difficulty
of the analysis was to cautiously remove the contribution of the varying
geocoronal \lya\ emission from the measurement.

The observation can be interpreted as the signature of an extended and escaping
exosphere of \ion{H}{i}. The authors argumentation is two-fold: (i) a Roche
lobe of radius $r_\perp \approx r_R = 2.8$~$r_p$ filled with \ion{H}{i} would
produce a dip of $\sim 11$\% in the transit light curve. The value measured is
larger, some hydrogen must be observed beyond the Roche lobe. Moreover, (ii)
the resolution of the spectrograph allowed the authors to report an absorption
spanning from -130 to +100~km~s$^{-1}$ with respect to the centre of the \lya\
stellar line. These velocities are above the escape velocity of the planet
(43~km~s$^{-1}$) and, therefore, hydrogen must again be escaping the planet.
These observations allowed to constrain the escape rate $\dot{m}$ of the planet
atmosphere to $\sim 10^{10}$~g~s$^{-1}$ (see \S\ref{sec:models}).

Ben-Jaffel (2007) measured an absorption of $(8.9\pm2.1)\%$, lower than the
value given by Vidal-Madjar \etal\ (2003). However, this difference is due to
the choice of a larger bandpass over the \lya\ line (see Fig.~\ref{fig:lya})
and the fact that the absorption takes place at the core of the line (see
Vidal-Madjar \etal\ 2008 for details).

\subsection{Constraints on the evaporation of HD\,209458b from unresolved data}
\begin{figure}
\begin{center}
\includegraphics[width=0.75\columnwidth]{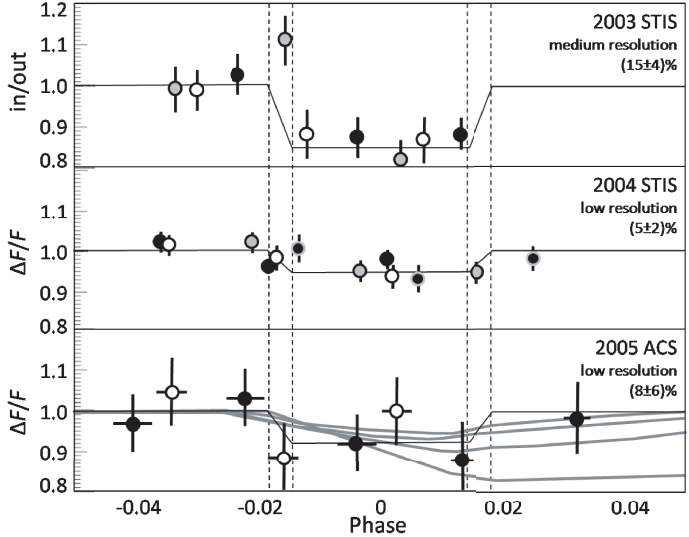}
\caption{\label{fig:light} Ultraviolet transit light curves of HD\,209458b at
\lya, for three different data sets. Different colours correspond to different
\emph{HST} visits. Trapezoidal fits to the light curves are included, whose
depths have been determined by (from top to bottom) Vidal-Madjar \etal\ (2003,
2004) and Ehrenreich \etal\ (2008). The bins used by these authors are
represented in Fig.~\ref{fig:lya}. Theoretical light curves resulting from the
passage of a hydrogen comet-like tail trailing the planet are figured in the
bottom panel, for different $\dot{m}$.}
\end{center}
\end{figure}

%------------------------------------------------------------------------------
The detection of HD\,209458b extended exosphere was first confirmed by
Vidal-Madjar \etal\ (2004) with low-resolution STIS spectra gathered during 4
transits of the planet. A low-resolution grism was employed to probe a large
spectral domain in the ultraviolet (from 120 to 170~nm) and test whether other
atomic elements could be seen in the exosphere. Light curves were calculated by
integrating over several unresolved atomic lines. Besides \ion{H}{i} \lya,
oxygen (\ion{O}{i}), carbon (\ion{C}{ii}), and various lines of silicium,
suflur, and nitrogen were tested. The 3 latter elements did not show any
significant absorption during the transit, whilst a $(5.3\pm1.9)\%$ dip was
observed in the \lya\ line (light curve in the middle panel of
Fig.~\ref{fig:light}). Because this value is obtained by integrating over the
entire unresolved \lya\ line, it is compatible with the $(15\pm4)\%$ previously
measured over $\sim1/3$ of the line.

Large absorptions of $(12.8\pm4.5)\%$ and $(7.5\pm3.6)\%$ during the transit
were also determined from fits to the light curve of the stellar flux
integrated over neutral oxygen (\ion{O}{i}, \ion{O}{i}$^{*}$, and
\ion{O}{i}$^{**}$) and ionized carbon (\ion{C}{ii} and \ion{C}{ii}$^{*}$)
lines, respectively.

This brought new constraints on the escape mechanism. In fact, atoms heavier
than hydrogen must be carried to high altitudes by the flow of escaping
hydrogen atoms with a velocity $u \sim 10$~km~s$^{-1}$ (at $10^4$~K), close to
the speed of sound: the escape is hydrodynamic, i.e., the escape rate is
enhanced by the upward bulk velocity at the top of the atmosphere, as in a
`blow-off' scenario (see \S\ref{sec:escape}). In addition, the observational
signatures of \ion{O}{i}$^*$, \ion{O}{i}$^{**}$, and \ion{C}{ii}$^*$ show that
collisions are necessarily happening at these high altitudes in order to
populate these excited levels. This implies Roche lobe densities of $n \sim
10^6$~cm$^{-3}$. It is remarkable that these simple observational constraints
provide an estimate of $\dot{m} = 4 \pi r_R^2 u \mu n / \mathrm{N_A} \sim
10^{10}$~g~s$^{-1}$, with $\mu$ the molar mass of H and $\mathrm{N_A}$ the
Avogadro number, which is compatible with previous, independent estimates (see
also Vidal-Madjar 2005).

\subsection{Are there other cases of evaporation?}
%-------------------------------------------------
The atmospheric evaporation phenomenon was first observed in 2003 when the
number of transiting planets around bright stars ($V\lesssim10$) was minute.
The situation drastically changed in 2004, when several `bright' transits were
spotted, in particular, HD\,149026b ($V = 8.15$; Sato \etal\ 2005) and
HD\,189733b ($V = 7.67$; Bouchy \etal\ 2005). Unfortunately, another drastic
event later this year was STIS stopping operations because of a power supply
failure. The Advanced Camera for Surveys (ACS; Ford \etal\ 2003) on board
\emph{HST} was then used as a back-up solution for programs aimed at obtaining
new data on these two planets in addition to HD\,209458b.

A prism can be inserted before the ACS detector to perform slitless
spectroscopy at \lya, with a spectral resolution close to STIS low-resolution
mode but -- as data revealed -- a lesser sensitivity. Transit light curves at
\lya\ can presently be obtained for the three above-mentioned planets. Archived
ACS data of HD\,149026b\footnote{Program \#10718, principal investigator: J.A.
Valenti.} seem not precise enough, though, to detect any potential extended
exosphere for this planet enriched in heavy element; this is mainly due to the
weak \lya\ emission of the transited G star.

Additional observations of HD\,209458b were obtained in 2005; their analysis
yielded a new measurement of the absorption of the exospheric hydrogen during
the transit: $(8.0\pm5.7)\%$ (Ehrenreich \etal\ 2008; light curve in the bottom
panel of Fig.~\ref{fig:light}). This result, although weakly significant,
remains compatible with previous detections, and acts as a benchmark for the
study of a more promising target: HD\,189733b. ACS data for this planet
orbiting a K star with a strong \lya\ emission are being obtained at the time
of writing.

\section{Models of atmospheric evaporation}
%==========================================
\label{sec:models}

\subsection{Interpreting the observations}
%-----------------------------------------
Models of extended exospheres have early been crafted by observers to justify
the search for large spectroscopic signatures (Rauer \etal\ 2000; Moutou \etal\
2001) and, later, to provide an interpretation of the $15\%$ absorption in the
\lya\ stellar line during the transit (Vidal-Madjar \etal\ 2003). These last
authors used a particle simulation in which the escape rate $\dot{m}$ is a free
parameter (Vidal-Madjar \& Lecavelier des Etangs 2004). Hydrogen atoms with
velocities $\sim v_\mathrm{esc}$ are blown by the planet at $\dot{m}$ and
sensitive to the known and wavelength-dependent stellar radiation pressure,
that accelerates them to the high velocities observed. The radiation pressure
on the moving planet carves the cloud of escaping atoms as a comet-like tail,
trailing the planet. Schneiter \etal\ (2007) employed a three-dimensional
hydrodynamical simulation to treat the interactions between the stellar wind
blown at a fixed rate and an isotropic wind of H atoms ejected by the planet
for various values of $\dot{m}$. Their simulations also show a cometary
structure for the cloud of escaping material.

Escaping atoms are also submitted to the planetary and stellar gravity.
Lecavelier des Etangs \etal\ (2004) considered the effect of tidal forces, and
found that tides shape the exosphere into an elongated `rugby
ball'\footnote{Roughly the same shape as an American football.}; they
calculated that the escape rate is enhanced toward the star and in the opposite
direction. The neutral H atoms are ionized by EUV photons on a short time scale
($\sim 6$~h). Thus, the size of the hydrogen tail should depend on the stellar
EUV flux.

The models then simulate the transit of the planet and its hydrogen cloud, and
convolve the resulting light curves with a synthetic \lya\ stellar profile.
They are able to reproduce the observed emission profile and the absorption
measured during the transit assuming $\dot{m} \sim 10^9$ to
$10^{10}$~g~s$^{-1}$. They also predict an asymmetric shape for the transit
curve, because of the passage of the comet-like tail following the planet.
Current observations do not have the precision nor the appropriate phase
coverage of the orbit to evidence this structure (Ehrenreich \etal\ 2008).

Holmstr\"om \etal\ (2008) proposed that the high velocities of the reported
absorption cannot be explained by the radiation pressure, but rather by the
recombination of solar wind protons in the planet exosphere \emph{via}
charge-exchange reactions (see \S\ref{sec:escape}). The cloud of neutral
hydrogen produced by their particle simulation is able to reproduce the
observed \lya\ profile during the transit, although it does so at the unlikely
condition that the radiation pressure is lower than the measured value and
independent of wavelength; they also need to assume extreme parameters for the
stellar wind -- a velocity of 50~km~s$^{-1}$ and a temperature of $10^6$~K at
0.04~\au. With these assumptions, they reproduced the \lya\ profile assuming an
escape rate decreased with respect to previous models, to $\sim
10^9$~g~s$^{-1}$.

\subsection{Understanding the dynamics of an evaporating atmosphere}
%-------------------------------------------------------------------
Numerical models of atmospheres enduring hydrodynamic escape were initially
built to explain the loss of hydrogen from the primitive terrestrial and
cytherean atmospheres (Watson, Donahue \& Walker 1981; Kasting \& Pollack 1983;
Chassefi\`ere 1996). The discovery of HD\,209458b evaporation sparked new
modelling insights, which led to a better understanding of the dynamics
subtending an energy-limited atmospheric escape. The evaporation was enhanced
in the past, as the EUV stellar irradiation is more intense for young stars
(Lammer \etal\ 2003; Penz, Micela \& Lammer 2008). This irradiation is
deposited across a large fraction of the extended atmosphere (Tian \etal\
2005), which mainly cools via adiabatic expansion and radiative emissions from
$\rm H_3^+$, a product of molecular hydrogen photoionization (Yelle 2004,
2006). Garc\'\i a-Mu\~noz (2007) provided one of the most exhaustive model of
HD\,209458b escaping atmosphere, including the effects of EUV heating, tidal
forces (also treated in Lecavelier des Etangs \etal\ 2004), chemistry and
ionization; this model is able to reproduce the observations of hydrogen,
carbon, and oxygen.

What all these detailed models basically agree on is that, despite some
possible one-to-two-orders-of-magnitude variations around the canonical value
of $\dot{m} \sim 10^{10}$~g~s$^{-1}$, the atmosphere of HD\,209458b -- and
probably of all massive hot Jupiters -- is stable to hydrodynamic escape. This
might not be the case, however, for lighter close-in exoplanets.

\section{Atmospheric evaporation in the frame of comparative exoplanetology}
%===========================================================================
\label{sec:exoplaneto}

Although a lot of detailed models has been developed to understand the physics
of evaporation, it appears that the resulting escape rate does not depend on
the details of the models but only on the assumed input energy used to escape
the potential well of the planet. As a consequence, an energy diagram was
introduced by Lecavelier des Etangs (2007) to evaluate the evaporation status
of exoplanets.

\begin{figure}
\begin{center}
\includegraphics[width=0.75\columnwidth]{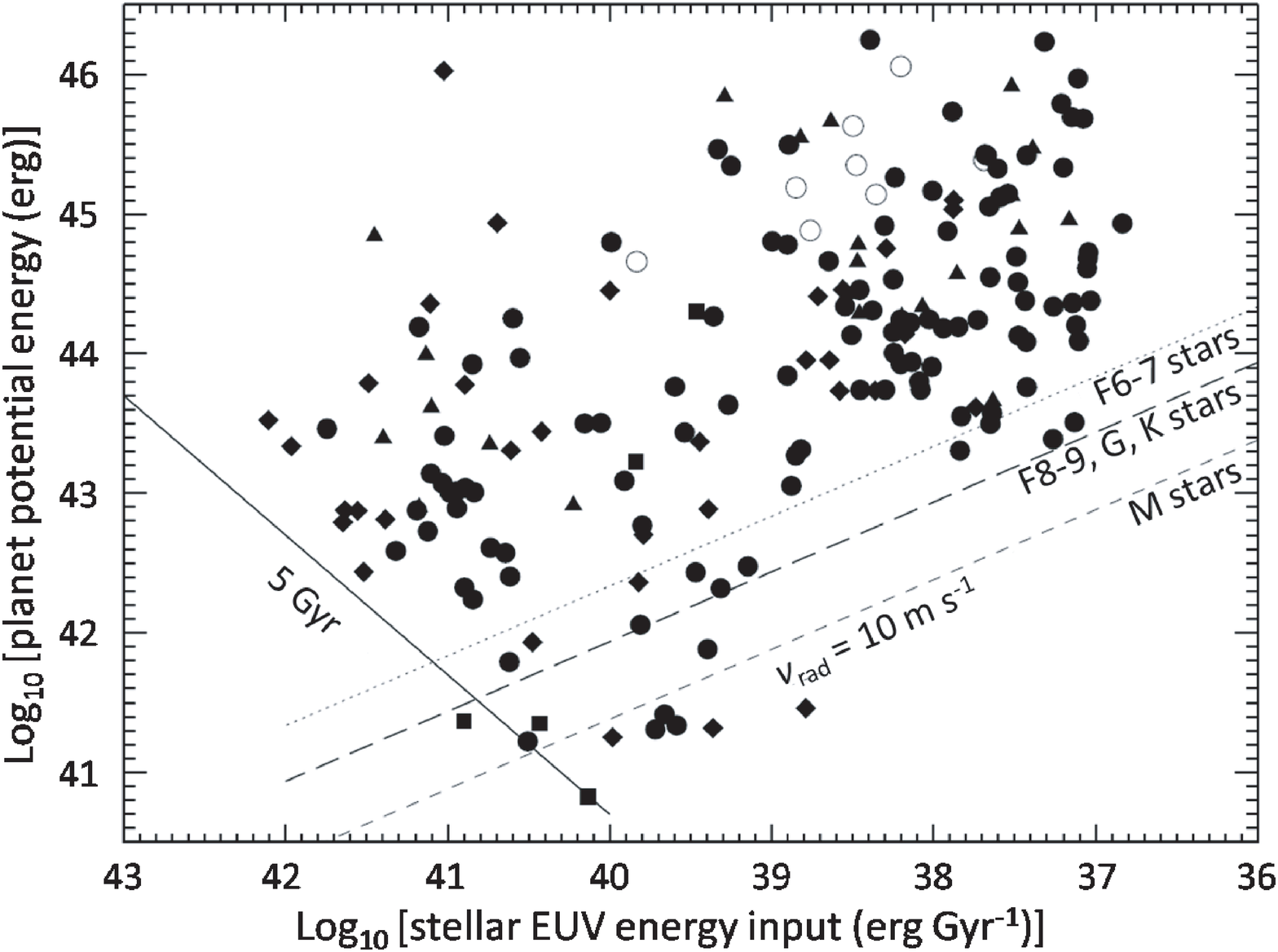}
\caption{\label{fig:diagram} Potential energy of exoplanets as a function of
the mean EUV energy received per billion years. The different symbols represent
planets around stellar types F ($\blacktriangle$), G ($\bullet$), K
($\blacklozenge$), and M ($\blacksquare$); a $\circ$ denotes a class-{\sc iii}
host star. The radial velocity detection limit of 10~m~s$^{-1}$ is plotted for
different types of stars (dotted, dashed, and long-dashed lines). The masses
and radii of non-transiting planets are taken to $\sqrt{2} m_p \sin i$ and
inferred from mass-radius curves of Guillot (2005), respectively. After
Lecavelier des Etangs (2007).}
\end{center}
\end{figure}

It is represented in Fig.~\ref{fig:diagram}, and consists in plotting the
potential energy one has to provide to a given planet in order to entirely
escape its atmosphere, as a function of the EUV energy available from the star
integrated through its life time. The ratio of these two quantities gives an
estimate of the planet life time. The lack of planets in the lower right part
of the diagram is a bias due to the precision of radial velocity measurements.
While no observational bias prevents planets to be found in the lower left-part
of the diagram (at low separations), there is clearly a dearth of planet in
this region. An enhanced evaporation could have make planets close to the 5-Gyr
isochrone, such as `hot-Neptunes' GJ\,436b (23~Earth masses, \mearth) and
GJ\,581b (16~\mearth), or `super-Earth' GJ\,876d (7.5~\mearth), evolved away
from the lower-left part of the diagram. These objects could be the remnants of
planets initially 3, 4.5, and 13 times more massive, respectively. Evaporation
should not have depleted GJ\,436b from a hydrogen envelope (Gillon \etal\
2007), hence it could still be evaporating today. Detecting its extended
exosphere and measuring its escape rate would bring new constraints on the fate
of close-in planets.

To test whether evaporation can alter a significant fraction of planets, Penz,
Micela \& Lammer (2008) estimated a distribution of stellar X-ray luminosity
and its impact on various distributions of planetary initial masses. They
showed that -- not surprisingly -- after 4~Gyr, 100\% of HD\,209458b-like
planets would survive to atmospheric escape. Barely less (95\%) would be wiped
out if they were twice closer to the star. In the meantime, 80\% of planets
with an initial mass and density (2~g~cm$^{-3}$) close to GJ~436b's would
resist. Interestingly, less than $50\%$ of lower-density hot-Neptunes ($\sim
1$~g~cm$^{-3}$) could survive evaporation, and 85\% could be eroded down to
super-Earth masses.

\section{Summary}
%================
A planetary transit in the ultraviolet produces remarkably large spectroscopic
signatures compared to what is observed in the visible. So far, the
observational results obtained on planet HD\,209458b have been obtained from
two different instruments and strengthened by independent data analysis. The
evaporation of close-in extrasolar planets also relies on a solid theoretical
ground. Progress will come from the observations of other cases of extended
exospheres and evaporation remnants. This will be feasible after \emph{HST}
Servicing Mission 4, which aims are -- among other few -- to fix STIS and
install the Cosmic Origins Spectrograph (COS; Green, Wilkinson \& Morse 2003).

\acknowledgements{The author warmly thanks A.~Vidal-Madjar and A.~Lecavelier
des Etangs for their careful readings and useful inputs. He acknowledges
support from the Agence Nationale de la Recherche through ANR grant
NT05-4\_44463.}

%-----------------------------
%      your bibliography
%-----------------------------

\end{document}